\documentclass[conference,a4paper,]{IEEEtran}

\usepackage{float}
\usepackage{amsmath}
\usepackage{amssymb}
\usepackage{graphicx}
\usepackage{esint}
\usepackage{epstopdf}
\usepackage[all]{xy}
\usepackage{balance}
\usepackage{psfrag}
\usepackage{subfig}
\usepackage{algorithm}
\usepackage{cite}
\usepackage{amsthm}

\makeatother

\begin{document}

\title{Iterative Detection for Compressive Sensing:\\ Turbo CS}

\author{\IEEEauthorblockN{Amin Movahed, Mark C. Reed}
\IEEEauthorblockA{School of Engineering and Information Technology, University of New South Wales, Canberra, Australia}
\IEEEauthorblockA{a.movahed@student.unsw.edu.au, mark.reed@unsw.edu.au}}

\maketitle

\begin{abstract}
We consider compressive sensing as a source coding method for signal transmission. We concatenate a convolutional coding system with 1-bit compressive sensing to obtain a serial concatenated system model for sparse signal transmission over an AWGN channel. The proposed source/channel decoder, which we refer to as turbo CS, is robust against channel noise and its signal reconstruction performance at the receiver increases considerably through iterations. We show 12 dB improvement with six turbo CS iterations compared to a non-iterative concatenated source/channel decoder.
\end{abstract}

\section{introduction}
In real transmission systems, source coding is used to minimize the transmitted bits. Moreover, channel coding is nearly always applied to minimize bit errors due to the channel noise. Therefore, source coding concatenated with channel coding is a recognized approach for reliable transmission of data \cite{hagenauer1995source}.

Compressive sensing (CS) is a new source coding approach in which signal measurement and compression are performed in a single step. The basic idea of CS is that any $ N $-dimensional signal which is $ K $-sparse (i.e., there are only $ K $ non-zero elements in the signal where $ N>>K $) is measured through few random linear projections. The sufficient number of projections, $ M $, guaranteeing signal reconstruction is often much less than $ N $ \cite{candes2005decoding}. Thus, CS can be considered as a method of data compression with rate $ N/M $. However, CS deals only with sparse or approximately sparse signals \cite{candes2008introduction}. In practice, many types of signals are sparse or can be represented with a sparse vector in a proper basis. Moreover, in some signal processing applications, e.g., magnetic resonance imaging, the processes of measurement and compression are not separable, and acquiring the signal through linear projections is an intrinsic part of the measuring process \cite{lustig2008compressed}.

In this paper, we use the principle of concatenated codes and turbo coding. Turbo codes are powerful channel encoding techniques first introduced by Berrou \textit{et al.} in 1993 \cite{berrou1993near} and the decoding performance achieves results close to the channel capacity. The encoding structure of a turbo encoder consists of a serial or parallel concatenation of convolutional encoders separated by random interleaving. 

In particular, we utilize the serial concatenated code approach\cite{benedetto1998serial}. 
The serial turbo decoder signal is decoded in an iterative process between two \textit{a posteriori probability} (APP) soft-input/soft-output decoders \cite{bahl1974optimal}.
 
The aim of this work is to apply a source encoder as the outer encoder concatenated with an inner channel decoder. In \cite{schmalen2011exit}, the authors introduced a turbo decoding approach by concatenating fixed length codes with convolutional codes for audio/video transmission. In this paper, we apply CS as a generic source coder for any kind of sparse signal. In order to do so, there are two main challenges:
\begin{itemize}
\item to input \textit{a posteriori} belief provided by the APP decoder to the CS decoder.
\item to calculate \textit{a priori} information from the CS decoder as input to the APP decoder for the next iteration.
\end{itemize}
As an approach, Bayesian CS \cite{ji2008bayesian}, which is a CS decoding method considering CS inversion from a Bayesian prospective, could be applied. Bayesian CS provides density function for each element of the reconstructed signal, which can be applied as \textit{a priori} information. However, the output of CS encoders is zero mean Gaussian distributed values while the input of convolutional encoders are $ -1 $ and $ +1 $. Thus, a special quantization is needed after a CS encoder.

In this work, we use 1-bit CS as the outer encoder. 1-bit CS is a quantized version of CS representing each measurement by only a two-state value\cite{boufounos20081}.

There are several methods introduced in the literature to solve 1-bit CS decoding problem. Some of these methods are based on linear and convex programming, e.g. \cite{CPA:CPA21442,plan2012robust}, while others are based on greedy methods \cite{boufounos20081,boufounos2009greedy,laska2011trust,jacques2011robust,kamilov2012one,yan2012robust,movahed2012robust}.  However, all the above mentioned methods only accept binary values as input to estimate the signal. In addition, none of these methods generates soft-valued \textit{a priori} information.

The key contribution of this paper is to propose a new reconstruction method for 1-bit CS which accepts soft-input and generates soft-output and, hence, is able to work iteratively together with an APP decoder to reconstruct the signal at receiver in the same fashion as in a classic serial concatenated turbo code.   

We refer to the proposed coding approach as turbo CS coding. The turbo CS encoder consists of the concatenation of a 1-bit CS encoder and a convolutional encoder at the transmitter. In the receiver, the turbo CS decoder iterates between an APP decoder and a 1-bit CS decoder.
Numerical experiments show a significant improvement in the quality of the reconstructed signal through turbo CS iterations.

\section{system model \label{setup}}
In this section, we describe the serial concatenated transmission and channel model. In the first part, we discuss 1-bit CS configuration and in the second part we combine 1-bit CS with a convolutional encoder.

\begin{figure*}[]
\centering
\psfrag{A}[][]{$\mathbf{x}$}
\psfrag{B}[][]{$\mathbf{y}$}
\psfrag{C}[][]{$\mathbf{b}$}
\psfrag{D}[][]{$\mathbf{d}$}
\psfrag{E}[][]{1-bit CS encoder}
\psfrag{F}[][]{$\mathbf{z}$}
\psfrag{CS encoder}[][]{CS encoder}
\psfrag{sign}[][]{$\textrm{sign}\left(\cdot\right)$}
\psfrag{Interleave}[][]{Interleaver}
\psfrag{Convolutional encoder}[][]{convolutional encoder}
\psfrag{AWGN}[][]{AWGN}
\psfrag{+}[][]{$+$}
\includegraphics[trim = 2mm 0mm 0mm 0mm,clip, scale=0.3]{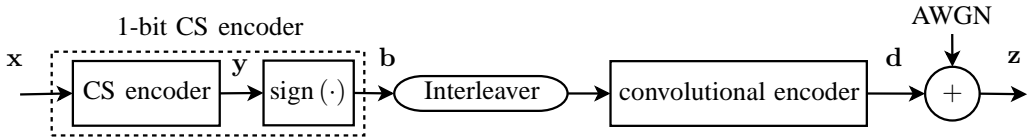}\caption{Transmitter model \label{trans}}
\end{figure*}

\subsection{1-bit compressive sensing}
In classic compressive sensing, each measurement $ y $ is obtained through a projection of $ K $-sparse signal, $ \mathbf{x}\in\mathbb{R}^N $, onto a random vector $ \boldsymbol{\phi}\in \mathbb{R}^N $. Therefore, for $ M $ number of measurements ($ M<N $) we have
\begin{equation}
\mathbf{y}=\boldsymbol{\Phi}\mathbf{x} \label{CS}
\end{equation}  
where $\boldsymbol{\phi}_{i}$ is the $ i $th row of $\boldsymbol{\Phi}\in\mathbb{R}^{M\times N}$ and $\mathbf{y}=\left[y_{1},y_{2},\ldots,y_{M}\right]^{\textrm{T}}$. It is shown that exact signal reconstruction is guaranteed when $ \boldsymbol{\Phi} $ satisfies the restricted isometry property \cite{candes2008introduction}. 

In most practical cases, obtained measurements need to be quantized before reconstruction. In the extreme case, which is referred to as 1-bit CS, measurements are represented by only one bit \cite{boufounos20081}. 1-bit CS output is essentially a sign function over CS measurements. Hence, binary measurements, $\mathbf{b}\in\left\{ -1,\,1\right\} ^{M}$, are obtained from
\begin{equation}
\mathbf{b}=\textrm{sign}\left(\mathbf{y}\right)=\textrm{sign}\left(\mathbf{\Phi x}\right)
 \end{equation}
where $\textrm{sign}\left(\cdot\right)$ denotes the sign function.

\subsection{Serially concatenated encoders}
At the transmitter, the interleaved binary output of the 1-bit CS encoder is encoded by a convolutional encoder. We denote the coded bits by $\mathbf{d}\in\{-1,+1\}^{P}$. $ M/P $ is the rate of the convolutional encoder. In turbo coding context, the 1-bit CS encoder and the convolutional encoder are referred to as outer and inner encoders respectively. The coded bits are transmitted through an AWGN channel with a known variance, $\sigma_{n}^{2}$. The channel output is then
\begin{equation}
\mathbf{z}=\mathbf{d}+\mathbf{n}
\end{equation}
where $\left[\mathbf{n}\right]_{i}\sim\textrm{N}\left(0,\sigma_{n}^{2}\right)$ and $\left[\cdot\right]_{i}$ denotes the $ i $th element in the argument. The system model is illustrated
in Fig. \ref{trans}.

In the next section, we propose an iterative method to reconstruct $ \mathbf{x} $ at the receiver from the noisy coded measurements $ \mathbf{z} $.
\section{Iterative 1-bit compressive sensing: Turbo CS\label{3}}
\subsection{A posteriori probability decoder\label{MAP}}
\textit{A posteriori probability} (APP) decoder is a soft-input/soft-output decoder \cite{bahl1974optimal}. APP takes two inputs: received signal $ \mathbf{z} $ and \textit{a priori} probability of elements of $ \mathbf{b} $ denoted by $ \boldsymbol{\alpha} $. Hence, we have
\begin{equation}
\mathbb{P}\left(\left[\mathbf{b}\right]_{i}=+1\right)=\left[\boldsymbol{\alpha}\right]_{i}\,\,\,\,\textrm{and}\,\,\,\,\, i=1,\ldots,M.
\end{equation} 
 
 At the output, APP gives \textit{a posteriori} probability of the elements of $ \mathbf{b} $ denoted by $ \boldsymbol{\alpha}^{\prime} $. Therefore,
 \begin{equation}
 \mathbb{P}\left(\left[\mathbf{b}\right]_{i}=+1|\mathbf{z}\right)=\left[\boldsymbol{\alpha}^{\prime}\right]_{i}\,\,\,\,\textrm{and}\,\,\,\,\, i=1,\ldots,M. \label{apostriori}
 \end{equation}
 Typically in a \textit{maximum a posteriori probability} decoder, a decision is made on $ \boldsymbol{\alpha^\prime} $ yielding hard-bits.

In iterative decoders, however, bit probabilities are exchanged between decoders, since they contain information about the reliability of the data. A vector containing soft-bits is denoted by $\mathbf{b}_{\textrm{soft}}\in\left\{ \left[-1,1\right]\right\} ^{M}$. Each element in  $\mathbf{b}_{\textrm{soft}}$ is defined as the expected value of the corresponding element in $ \mathbf{b} $. Hence, for \textit{a priori} soft-bits we have
\begin{eqnarray}
 & \left[\mathbf{b}_{\textrm{soft}}\right]_{i} & =\mathbb{E}\left(\left[\mathbf{b}\right]_{i}\right)=\mathbb{P}\left(\left[\mathbf{b}\right]_{i}=+1\right)-\mathbb{P}\left(\left[\mathbf{b}\right]_{i}=-1\right)\nonumber\\
 &  & =\left[\boldsymbol{\alpha}-\left(1-\boldsymbol{\alpha}\right)\right]_{i}=2\left[\boldsymbol{\alpha}\right]_{i}-1.
\end{eqnarray}   
In the same way, \textit{a posteriori} soft-bits are obtained from
\begin{eqnarray}
 & \left[\mathbf{b}_{\textrm{soft}}^{\prime}\right]_{i} & =\mathbb{E}\left(\left[\mathbf{b}\right]_{i}|\mathbf{z}\right)=2\left[\boldsymbol{\alpha}^{\prime}\right]_{i}-1. \label{bsoft}
\end{eqnarray} 
Furthermore, hard-bits are denoted with $ \mathbf{b}_{\textrm{hard}} $ and we have
\begin{equation}
\mathbf{b}_{\textrm{hard}}=\textrm{sign}\left(\mathbf{b}_{\textrm{soft}}^{\prime}\right)
\label{bhard}.
\end{equation}
Intuitively, when $ \mathbb{P}\left(\left[\mathbf{b}\right]_{i}=+1\right) $ is $ 0 $, the $ i $th soft-bit is $ -1 $ and when $ \mathbb{P}\left(\left[\mathbf{b}\right]_{i}=+1\right) $ is $ 1 $,  the $ i $th soft-bit is  $ +1 $.

In iterative decoding, the inner decoder needs to receive the parameter $ \mathbf{b}_\textrm{soft} $ and estimate $ \mathbf{x} $ and $ \mathbf{b}_\textrm{soft}^{\prime} $. In the next two sections, we give a brief review on 1-bit CS reconstruction and then introduce a 1-bit CS algorithm that can be used in an iterative turbo CS decoder where the CS constituent decoder accepts soft bits in and generates soft bits out.
\subsection{1-bit CS reconstruction algorithm\label{1bit}}
The aim of a 1-bit CS reconstruction algorithm is to estimate the values in a vector $ \mathbf{x} $ based on an observation vector $ \mathbf{b} $ and knowing the measuring matrix $ \boldsymbol{\Phi} $. In many practical cases, there might be some random bit flips in $ \mathbf{b} $ due to the quantization error or noise in the transmission process. The number of these bit flips is a measure of the noise level. Some of the reconstruction algorithms consider the number of the bit flips to reconstruct the signal efficiently and are robust against the random bit flips in the binary measurements \cite{yan2012robust,movahed2012robust}. 

Among all 1-bit CS reconstruction algorithms, \textit{adaptive outlier pursuit with bit flips} (AOP-f) \cite{yan2012robust} has the best reconstruction performance in the presence of random bit flips and when the sparsity level of the signal and the number of the bit flips are known. There are two types of AOP-f based on $ \ell_1 $-norm minimization (AOP-$ \ell_1 $-f) and $ \ell_2 $-norm minimization (AOP-$ \ell_2 $-f). Since AOP-$ \ell_1 $-f outperforms AOP-$ \ell_2 $-f in terms of signal reconstruction performance, we focus on AOP-$ \ell_1 $-f in this paper. Henceforth, we refer to AOP-$ \ell_1 $-f as AOP-f. 

AOP-f is an iterative algorithm that estimates $\mathbf{x}$ and the position of the bit flips in $ \mathbf{b} $. $\tilde{\mathbf{b}}$ denotes the noisy binary measurements vector and $ L $ denotes the number of the bit flips in $\tilde{\mathbf{b}}$. The position of the random bit flips in $ \tilde{\mathbf{b}} $ is represented by vector $ \boldsymbol{\Omega}\in\left\{ -1,\,1\right\} ^{M}$ where $\boldsymbol{\Omega}=\mathbf{b}\odot\tilde{\mathbf{b}}$ and $ \odot $ denotes element-wise product. That is, $\left[\boldsymbol{\Omega}\right]_{i}=-1 $ means that there is a bit flip in $\left[\tilde{\mathbf{b}}\right]_{i}$. AOP-f solves the following optimization problem
\begin{eqnarray}
 &  & \left(\hat{\mathbf{x}},\hat{\mathbf{\Omega}}\right)=\underset{\mathbf{x},\mathbf{\Omega}}{\arg\min}\left\Vert \left(\tilde{\mathbf{b}}\odot\mathbf{\Omega}\odot\Phi\mathbf{x}\right)^{-}\right\Vert _{1}\nonumber \\
 &  & \,\qquad\mathrm{s.t.}\;\frac{1}{2}\sum_{i}\left(1-\left[\mathbf{\Omega}\right]_{i}\right)\leq L\nonumber \\
 &  & \,\;\;\qquad\quad\left\Vert \mathbf{x}\right\Vert _{0}\leq K\nonumber \\
 &  & \,\,\,\,\quad\qquad\left\Vert \mathbf{x}\right\Vert _{2}=1\
\label{AOPf}
\end{eqnarray}
where 
$\left\Vert \cdot\right\Vert _{p}$ denotes $ \ell_p $-norm\footnote{$\left\Vert \mathbf{x}\right\Vert _{p}:=\left(\sum_{i=1}^{N}\left|\left[\mathbf{x}\right]_{i}\right|^{p}\right)^{1/p}$} of the argument and  $\left(\cdot\right)^{-}$ is negative function defined as
\[
\left(\left[\mathbf{x}\right]_{i}\right)^{-}=\begin{cases}
\left|\left[\mathbf{x}\right]_{i}\right|, & \textrm{if\,\,}\left[\mathbf{x}\right]_{i}<0,\\
0, & \textrm{otherwise.}
\end{cases}
\]

In the next section, we propose some changes to the input of AOP-f to be able to utilize soft-bits as input. In addition, we apply a mapping method on the reconstructed signal to produce \textit{a priori} soft-bits to be used as an input to the APP decoder. 

\subsection{Soft-in/soft-out 1-bit CS decoder \label{sec2}}
As mentioned in section \ref{1bit}, AOP-f accepts binary values as input to reconstruct the signal. Therefore, a trivial way to apply AOP-f as a decoder after the APP decoder is to use $ \mathbf{b}_{\textrm{hard}} $ from (\ref{bhard}) in (\ref{AOPf}). However, by solely using hard-bits, we lose information about the reliability of the data. In addition, AOP-f needs to know an estimate of the number of the bit flips in $ \mathbf{b}_{\textrm{hard}} $ to reconstruct the signal efficiently. 

Here, we develop a method to use soft-bits as input to reconstruct the signal via AOP-f.
$ \tilde{\mathbf{b}} $ is replaced with $ \mathbf{b}_{\textrm{hard}} $ in (\ref{AOPf}). In addition, we define $ \boldsymbol{\alpha}_{\textrm{flip}} $ whose elements represent the probability of a bit flip in the corresponding element of $ \mathbf{b}_\textrm{hard} $. Thus, $ \boldsymbol{\alpha}_{\textrm{flip}} $ is derived from
\begin{equation}
\left[\boldsymbol{\alpha}_{\textrm{flip}}\right]_{i}=\begin{cases}
\mathbb{P}\left(\left[\mathbf{b}\right]_{i}=-1|\mathbf{z}\right), & \textrm{if\,\,}\left[\mathbf{b}_{\textrm{hard}}\right]_{i}=1,\\
\mathbb{P}\left(\left[\mathbf{b}\right]_{i}=+1|\mathbf{z}\right), & \left[\mathbf{b}_{\textrm{hard}}\right]_{i}=-1.
\end{cases} \label{pflip}
\end{equation}
Substituting (\ref{apostriori}), (\ref{bsoft}) and (\ref{bhard}) in (\ref{pflip}) gives
\begin{equation}
\left[\boldsymbol{\alpha}_{\textrm{flip}}\right]_{i}=\begin{cases}
1-\left[\boldsymbol{\alpha}^{\prime}\right]_{i}, & \textrm{if\,\,}\left[\boldsymbol{\alpha}^{\prime}\right]_{i}\geq0.5,\\
\left[\boldsymbol{\alpha}^{\prime}\right]_{i}, & \textrm{otherwise.}
\end{cases}
\end{equation}

The estimated number of the bit flips is denoted by $ \bar{L} $ and is obtained from
\begin{equation}
\bar{L}=\textrm{round}\left(\sum_{i=1}^{M}\left[\boldsymbol{\alpha}_{\textrm{flip}}\right]_{i}\right).
\label{L}
\end{equation}
Now with $\mathbf{b}_{\textrm{soft}}^{\prime}$ from (\ref{bsoft}) and $ \bar{L} $ from (\ref{L}), $ \mathbf{x} $ can be estimated through AOP-f and the following optimization can be solved via the algorithm in  \cite{yan2012robust}
\begin{eqnarray}
 &  & \left(\hat{\mathbf{x}},\hat{\mathbf{\Omega}}\right)=\underset{\mathbf{x},\mathbf{\Omega}}{\arg\min}\left\Vert \left(\textrm{sign}\left(\mathbf{b}_{\textrm{soft}}^{\prime}\right)\odot\mathbf{\Omega}\odot\Phi\mathbf{x}\right)^{-}\right\Vert _{1}\nonumber \\
 &  & \,\qquad\mathrm{s.t.}\;\frac{1}{2}\sum_{i}\left(1-\left[\mathbf{\Omega}\right]_{i}\right)\leq \bar{L}\nonumber \\
 &  & \,\;\;\qquad\quad\left\Vert \mathbf{x}\right\Vert _{0}\leq K\nonumber \\
 &  & \,\,\,\,\quad\qquad\left\Vert \mathbf{x}\right\Vert _{2}=1\ \label{AOPfs}.
\end{eqnarray}

The next step of the decoder generates soft-bits, $ \mathbf{b}_{\textrm{soft}} $, at the output. We apply a CS encoder over the estimated signal. Thus, we obtain
\begin{equation}
\mathbf{y}^{\prime}=\boldsymbol{\Phi}\hat{\mathbf{x}}.
\end{equation}

Elements of $ \mathbf{y}^{\prime} $ can be approximated by a Gaussian distribution with zero mean. In this case, unlike \textit{binary phase shift keying} (BPSK) system, most of the received values to be mapped are concentrated around $ 0 $. The challenge is to map these values to an interval between $ -1 $ and $ 1 $ based on their reliabilities. The elements with values around $ 0 $ are the least reliable for generating \textit{a priori} soft-values. The elements with the most reliability are the ones that are the furthest from $ 0 $. Therefore, we utilize elements of $ \mathbf{y}^{\prime} $ that are further from $ 0 $, and over iterations, we consider the influence of the elements of $ \mathbf{y}^{\prime} $ with values closer and closer to zero. 

In the case that either there is no noise in the received binary measurements or the estimation of the number of the bit flips is exact, $ \mathbf{y}^{\prime} $ is very close to $ \mathbf{y} $ and the sign of each element of $ \mathbf{y}^{\prime} $ describes the sign of the corresponding element in  $ \mathbf{b} $. In the noisy case, however, there are some sign mismatches between the elements of $ \mathbf{y}^{\prime} $ and $  \mathbf{y} $. To consider the effect of the random bit flips on the soft-values, we multiply $ \mathbf{b}_{\textrm{hard}} $ with $ \mathbf{y}^{\prime} $ and the result is denoted by $ \boldsymbol{\psi} $, 
\begin{equation}
\boldsymbol{\psi}=\textrm{sign}(\mathbf{b}^\prime_{\textrm{soft}})\odot\mathbf{y}^{\prime}=\mathbf{b}_{\textrm{hard}}\odot\mathbf{y}^{\prime}.
\label{psi}
\end{equation}
 
In fact, the element-wise multiplication in (\ref{psi}) removes the sign of the elements of $ \mathbf{y}^{\prime} $. In the case that there is no bit flips in $ \mathbf{b}_{\textrm{hard}} $, then $\mathbf{b}_{\textrm{hard}}=\mathbf{b}=\textrm{sign}\left(\mathbf{y}^{\prime}\right)$ and all the elements of (\ref{psi}) are positive. However, in the presence of the random bit flips, the negative elements of $ \boldsymbol{\psi} $ depict the sign flips in $ \mathbf{b}_{\textrm{hard}} $ and the elements with large amplitudes are more reliable than the ones with small and negative amplitudes. Based on the above facts, a mapping function is introduced which maps each element of $ \boldsymbol{\psi} $ to a real value between $ -1 $ and $ 1 $. The mapping function $ \Lambda\left(\boldsymbol{\psi}\right) $ is defined as follows 
\begin{equation}
\Lambda\left(\boldsymbol{\psi}\right)=\begin{cases}
\textrm{min}\left(1,\frac{\left[\boldsymbol{\psi}\right]_{i}}{\gamma\cdot\textrm{max}\left(\boldsymbol{\psi}\right)}\right), & \textrm{if\,\,}\left[\boldsymbol{\psi}\right]_{i}\geq0,\\
\frac{\left[\boldsymbol{\psi}\right]_{i}}{\left|\textrm{min}\left(\boldsymbol{\psi}\right)\right|}, & \left[\boldsymbol{\psi}\right]_{i}<0
\label{map}
\end{cases}
\end{equation}
where $0\leq\gamma\leq1$ is the normalized Euclidean distance between $ \mathbf{b}^{\prime}_{\textrm{soft}} $ and $ \mathbf{b}_{\textrm{hard}} $. We have  
\begin{equation}
\gamma=\frac{\left\Vert \mathbf{b}_{\textrm{soft}}^{\prime}-\textrm{sign}\left(\mathbf{b}_{\textrm{soft}}^{\prime}\right)\right\Vert _{2}}{\sqrt{M}}=\frac{\left\Vert \mathbf{b}_{\textrm{soft}}^{\prime}-\mathbf{b}_{\textrm{hard}}\right\Vert _{2}}{\sqrt{M}} \label{gamma}.
\end{equation}
In fact, $ \gamma $ determines how much information is lost by applying sign function over $ \mathbf{b}^{\prime}_{\textrm{soft}} $.

Since the signs of the elements in $ \mathbf{y}^{\prime} $ were removed in (\ref{psi}), the obtained values from (\ref{map}) need to be multiplied again by $ \mathbf{b}_{\textrm{hard}} $ in order to bring the signs back. Hence, the soft-output is obtained by 
\begin{equation}
\mathbf{b}_{\textrm{soft}}=\Lambda\left(\boldsymbol{\psi}\right)\odot\mathbf{b}_{\textrm{hard}}
\label{bsoft2}.
\end{equation}

\begin{figure}[t]
\vspace{0.3mm}

\centering

\psfrag{A}[][]{$\scriptstyle \textrm{min}\left(\boldsymbol{\psi}\right)$}
\psfrag{B}[][]{$\scriptstyle \gamma\cdot\textrm{max}\left(\boldsymbol{\psi}\right)$}
\psfrag{C}[][]{$\scriptstyle \textrm{max}\left(\boldsymbol{\psi}\right)$}
\psfrag{D}[][]{$\scriptstyle \textrm{PDF} $}
\psfrag{E}[][]{$\scriptstyle \left[\boldsymbol{\psi}\right]_{i}$}
\psfrag{F}[][]{$\scriptstyle \left[\Lambda\left(\boldsymbol{\psi}\right)\right]_{i}$}
\psfrag{G}[][]{$\scriptstyle \left[\boldsymbol{\psi}\right]_{i}$}
\psfrag{H}[][]{$\scriptstyle \left[\Lambda\left(\boldsymbol{\psi}\right)\right]_{i}$}
\psfrag{I}[][]{$\scriptstyle \left[\mathbf{b}_{\textrm{soft}}\right]_{i} $}
\psfrag{J}[][]{$\scriptstyle \textrm{PDF}$}
\psfrag{K}[][]{$\scriptstyle \textrm{PDF}$}
\psfrag{L}[][]{$\scriptstyle  +1 $}
\psfrag{M}[][]{$\scriptstyle -1 $}
\psfrag{Z}[][]{$\scriptstyle 0 $}

\includegraphics[trim = 2mm 0mm 0mm 0mm,clip, scale=0.575]{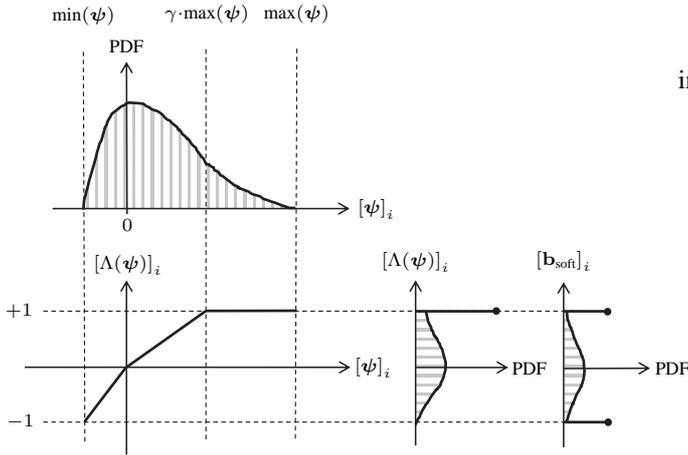}\caption{Mapping method \label{diag}}
\end{figure}
In Fig. \ref{diag}, the mapping method is depicted. In words, $ \Lambda\left(\boldsymbol{\psi}\right) $ is a mapping function that categorizes the elements of $ \boldsymbol{\psi} $ by their signs: 
\begin{itemize}
\item The negative elements of $ \boldsymbol{\psi} $ are mapped to values in an interval between $ -1 $ and $ 0 $ based on their amplitudes. As mentioned above, the negative elements in $ \boldsymbol{\psi} $ specify the bit flips in $ \mathbf{b}_{\textrm{hard}} $. In addition, the negative elements with small values are more likely to be flipped and are mapped to values close to $ -1 $.

\item The positive elements of $ \boldsymbol{\psi} $ are mapped based on their amplitudes between $ 0 $ and $ \gamma\cdot\textrm{max}\left(\boldsymbol{\psi}\right)$ to values between $ 0 $ and $ +1 $. Elements of $ \boldsymbol{\psi} $ exceeding $ \gamma\cdot\textrm{max}\left(\boldsymbol{\psi}\right)$ are clipped and mapped to $ +1 $.

\end{itemize}

We refer to the proposed decoding method as soft-in/soft-out 1-bit CS decoder.

\noindent \textit{Example:} To justify the performance of the soft-in/soft-out 1-bit CS decoder, we consider the best case where there is no noise in the binary measurements. Hence, 
$\left[\boldsymbol{\alpha}_{\textrm{bit}}\right]_{i}=0\,\,\,\, \textrm{for}\,\,\,\,i=1\ldots\, M $.
We have $ \bar{L}=0 $ from (\ref{L}). $ \hat{\mathbf{x}} $ is estimated by (\ref{AOPfs}). Elements of $ \boldsymbol{\psi} $ obtained from (\ref{psi}) are all positive values. Therefore, $\textrm{min}\left(\boldsymbol{\psi}\right)=0$. Furthermore, $\textrm{sign}\left(\mathbf{b}_{\textrm{soft}}^{\prime}\right)=\mathbf{b}_{\textrm{hard}}$ and (\ref{gamma}) gives $ \gamma =0$ that yields $\gamma\cdot\textrm{max}\left(\boldsymbol{\psi}\right)=0$. Thus, all the elements of $ \Lambda\left(\boldsymbol{\psi}\right) $ are $ 1 $. In this case, $ \mathbf{b}_{\textrm{soft}} $, given by (\ref{bsoft2}), is identical to $ \mathbf{b}^{\prime}_{\textrm{soft}} $. 

\subsection{Combination of soft-in/soft-out 1-bit CS and APP decoding}
In section \ref{sec2}, the soft-in/soft-out 1-bit CS reconstruction method was introduced which receives soft-bits and generates improved soft-bits as output. In this section, we combine the soft-in/soft-out 1-bit CS decoder with an APP decoder to obtain the turbo CS decoder for the transmission system in section \ref{setup}. 

\begin{figure}[t]
\hspace{-0.5cm}
\psfrag{A}[][]{$ \mathbf{y}^{\prime} $}
\psfrag{B}[][]{$ \mathbf{b}_{\textrm{hard}} $}
\psfrag{C}[][]{$ \mathbf{b}_{\textrm{soft}}^{\prime} $}
\psfrag{D}[][]{$ \mathbf{b}_{\textrm{soft}} $}
\psfrag{E}[][]{$ \mathbf{z} $}
\psfrag{F}[][]{$ \hat{\mathbf{x}} $}
\psfrag{Interleave}[][]{Interleave}
\psfrag{De-Interleave}[][]{De-Interleave}
\psfrag{APP decoder}[][]{APP decoder}
\psfrag{Mapping}[][]{Mapping}
\psfrag{sign(.)}[][]{$\textrm{sign}\left(\cdot\right)$}
\psfrag{1-bit CS decoder}[][]{1-bit CS decoder}
\psfrag{A priori}[][]{a priori}
\psfrag{Information}[][]{information}
\psfrag{A posteriori}[][]{a posteriori}
\psfrag{bit flips}[][]{bit flips}
\psfrag{estimator}[][]{estimator}
\psfrag{ Probability}[][]{ probability}
\psfrag{CS encoder}[][]{CS encoder}
\psfrag{soft-in/soft-out 1-bit CS decoder}[][]{soft-in/soft-out 1-bit CS decoder}
\includegraphics[trim = 0mm 0mm 0mm 0mm,  width=9cm]{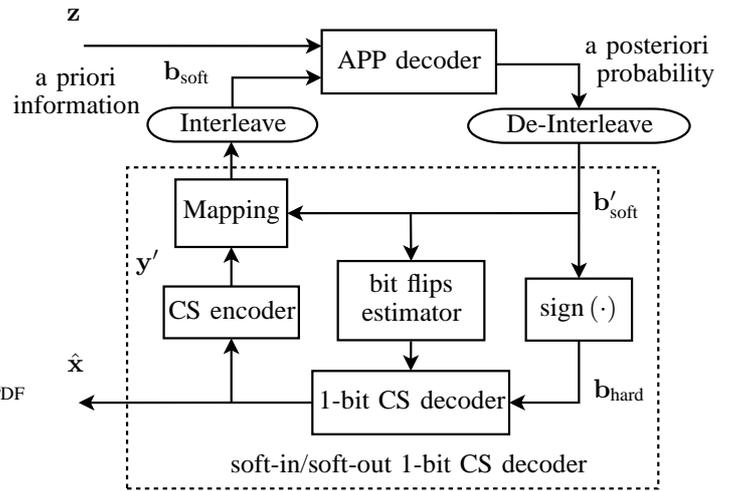}\caption{Turbo CS decoder \label{turbo}}
\end{figure}

As discussed in section \ref{setup}, the transmission system consists of a 1-bit CS encoder serially concatenated with a convolutional encoder at the transmitter. Hence, the 1-bit CS encoder works as a source encoder that receives real values and compresses the data with rate $ M/N $. The binary output of the 1-bit CS encoder is given to the convolutional encoder.  At the receiver, as illustrated in Fig. \ref{turbo}, the received noisy signal is input to an APP decoder. The \textit{a priori} soft-bits are zero for the first iteration. The soft-output of the decoder, namely \textit{a posteriori} probability, is given to the soft-in/soft-out 1-bit CS decoder to estimate the transmitted signal. The soft-output of the soft-in/soft-out 1-bit CS decoder is provided to the APP decoder as \textit{a priori} information for the next iteration. These steps are repeated for each iteration. Through the iterations and as $ \mathbf{b}_{\textrm{soft}}^{\prime} $ tends to $ \mathbf{b} $, $ \gamma $ goes to $ 0 $ and the output of the turbo CS decoder converges.
\vspace{10mm}
\begin{figure*}[]
\begin{center}
\hspace{-7mm}
\psfrag{A}[c][b]{$\textrm{E}_{\textrm{b}}/\textrm{N}_{0}$ (dB)}
\psfrag{B}[c][b]{RSNR (dB)}
\psfrag{C                                        }[l][]{Iteration 1}
\psfrag{D}[l][]{Iteration 2}
\psfrag{E}[l][]{Iteration 3}
\psfrag{F}[l][]{Iteration 4}
\psfrag{G}[l][]{Iteration 5}
\psfrag{H}[l][]{Iteration 6}
\psfrag{I}[l][]{uncoded 1-bit CS}
\psfrag{J}[l][]{12 dB}
\includegraphics[trim = 0mm 0mm 0mm 0mm,  width=2.1\columnwidth]{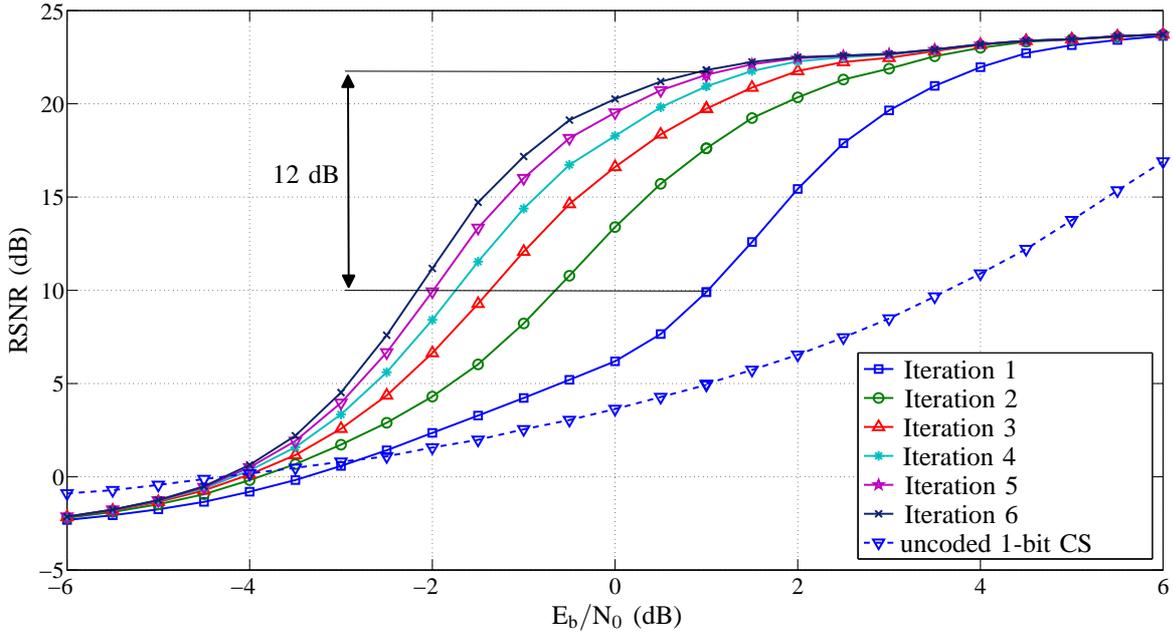}\caption{Turbo CS reconstruction performance \label{graph}}
\end{center}
\end{figure*}
\vspace{-0.7cm}
\section{Numerical results}
In this section, we verify the reconstruction performance of turbo CS through numerical simulation. We choose $ K $-sparse signal vector $ \mathbf{x} $ randomly in each realization. We set the dimension of the signal $ N=1000 $ and its sparsity level $ K=10 $. The non-zero elements of $ \mathbf{x} $ follow zero-mean Gaussian distribution with variance $ 1 $. These elements are distributed uniformly through the signal vector $ \mathbf{x} $. The elements of measuring matrix $ \boldsymbol{\Phi} $ are generated based on a Gaussian distribution with zero mean and variance $ 1/M $. The number of the encoded bits is set to $ M=500 $. Thus, the rate of the 1-bit CS encoder is $ N/M=2 $. The signal is encoded through the 1-bit CS encoder and its binary output is interleaved by a random interleaver with block length $ 500 $. However, simulation results show that the reconstruction performance of the turbo CS decoding system is not sensitive to the interleaver block length. 

The interleaved bits are passed to a G[5,7] convolutional encoder with memory=$ 2 $, four states and rate $ M/P=1/2 $. Then, the output of the convolutional encoder is passed through an AWGN channel with noise variance $ \sigma^2_n $. We show the power of the channel noise by signal to noise ratio (SNR) which is defined as  
\begin{equation}
\textrm{SNR}=\frac{\textrm{E}_{\textrm{b}}}{\textrm{N}_{0}}=\frac{1}{2R\sigma_{n}^{2}} \label{SNR}
\end{equation}
where $ \textrm{E}_{\textrm{b}} $ denotes the averaged power of a bit at the input of the channel encoder and $ R $ denotes the encoder rate which is $ 1/2 $ for G[5,7]. 

The channel output is decoded by our proposed turbo CS decoder. To show the reconstruction performance, received signal to noise ratio (RSNR) is defined as follows
\begin{equation}
\textrm{RSNR}=\frac{\mathbb{E}\left(\left\Vert \mathbf{x}\right\Vert _{2}^{2}\right)}{\mathbb{E}\left(\left\Vert \mathbf{x}-\hat{\mathbf{x}}\right\Vert _{2}^{2}\right)}.
\end{equation}

We verify the reconstruction performance of turbo CS through iterations in different channel noise scenarios. The signal to noise ratio is varied between $ -6 $ dB and $ 6 $ dB and the calculated RSNR is averaged over $ 10^4 $ realizations. Simulated results are shown in Fig. \ref{graph} with $ 1 $ to $ 6 $ iterations of the turbo CS decoder. 

As it can be seen in Fig. \ref{graph}, there is a huge improvement in the reconstruction performance of turbo CS through iterations. The reconstruction performance converges after around six iterations. We achieve $ 12 $ dB of improvement at $\textrm{E}_{\textrm{b}}/\textrm{N}_{0}=1$ dB. This is a massive performance gain over concatenated coding with no iterations (iteration 1 in Fig. \ref{graph}). Note we see the turbo like properties where most of the gain ($ 7.5 $ dB) comes in the 2nd iteration. After convergence, the difference between the reconstruction accuracy of turbo CS when the channel is very noisy ($ \frac{\textrm{E}_{\textrm{b}}}{\textrm{N}_{0}} = 1 $ dB) and when the channel is almost noiseless ($ \frac{\textrm{E}_{\textrm{b}}}{\textrm{N}_{0}} = 6 $ dB) is just around $ 2 $ dB. 

In another simulation, the convolutional encoder is removed. In this case, the channel noise is calculated by (\ref{SNR}) where $ R=1 $. Since there is no information at the receiver about the number of the random bit flips in the received signal, we set $ \bar{L}=0 $ in (\ref{AOPfs}). The performance of uncoded 1-bit CS is depicted by dashed line in Fig. \ref{graph}. It can be seen that RSNR of 1-bit CS decoding is significantly worse when there is no channel encoding/decoding used. 

Note that when SNR is less than $ -4 $ dB, uncoded 1-bit CS outperforms turbo CS. This behaviour is not unexpected since in general when the AWGN channel is very noisy, convolutional decoders have poor performance in terms of bit error rate in comparison to an uncoded BPSK system \cite{proakisdigital}.

\section{Conclusion}

In this work, we applied 1-bit CS as a generic source encoding method in a signal transmission problem over an AWGN channel. We combined 1-bit CS with a convolutional encoder and formed a serial concatenated source/channel encoding method. The key contribution of this paper is the turbo CS decoding method for the above transmission system. In turbo CS, we benefit from \textit{a posteriori} soft-bits generated by the APP decoder to estimate the reliability (number of the sign flips) of the bits given to the 1-bit CS decoder. In addition, a mapping method was introduced to modify the given soft-bits based on the current estimation of the signal.

Here, we used a non-recursive Convolutional Code G[5,7] as the channel encoder and the appropriate APP decoder within our turbo CS decoder. However, we expect that most convolutional endcoder/decoder could be applied to this system model to reconstruct the signal jointly with the soft-in/soft-out 1-bit CS decoder. In addition, unlike classic turbo coding, turbo CS performance is not sensitive to the length of the interleaver. 

Simulation results show that the reconstruction performance of turbo CS improves considerably through iterations. When the channel is very noisy (SNR=$ 1 $ dB) $ 12 $ dB gain is achievable after six iterations. In addition, the performance of the converged turbo CS is robust against the channel noise.


\balance
\bibliographystyle{IEEEtran}

\bibliography{IEEEabrv,reference}

\end{document}